\documentclass[aps,twocolumn, pra]{revtex4}
\usepackage{newlfont}
\usepackage{amssymb}
\usepackage{amsfonts}
\usepackage{amsmath}
\usepackage{bm}

\usepackage{epsfig}
\usepackage{newlfont}
\usepackage{amssymb}
\usepackage{amsfonts}
\usepackage{amsmath}
\usepackage{bm}

\begin{document}

\title{Geometry versus Entanglement in Resonating Valence Bond Liquids}

\author{Himadri Shekhar Dhar\(^{1}\) and Aditi Sen(De)\(^{2,1}\) }

\affiliation{\(^1\)School of Physical Sciences, Jawaharlal Nehru University, New Delhi 110067, India\\
\(^2\)Harish-Chandra Research Institute, Chhatnag Road, Jhunsi, Allahabad 211019, India
}

\begin{abstract}

%Importance of resonating valence bond states  comes from its relation to high temperature superconductivity. 
%It was shown that even pseudo-two dimensional system like ladder for some metal can show its ground state as resonating 
%valence bond state. 

We  investigate the behavior of bipartite as well as genuine multipartite entanglement of a resonating valence bond state on a ladder. 
We show that the system possesses significant amounts of bipartite entanglement in the steps of the ladder while no substantial bipartite entanglement 
is present in the rails. Genuine multipartite entanglement present in the system is negligible. The results are in stark contrast 
with the entanglement properties of the same state on isotropic lattices in two and higher dimensions, indicating that the geometry of the lattice can have 
important implications on the quality of quantum information and other tasks that can be performed by using multiparty states on that lattice.

%Our results shows the substantial
% change in both bipartite as well as multipartite entanglement in comparison with the resonating valence bond states in a square or higher-dimensional lattice. 

%The study aims at understanding the entanglement properties of Resonating Valence Bond (RVB)states on a quasi-two dimensional (ladder) system. It has been established that in RVB states of two and higher dimension there is negligible amount of bipartite entanglement though the system as a whole is multi-party entangled. We investigate this property in the quasi-two dimensional RVB system and try understand the amount of bipartite entanglement present for various sizes of the lattice and also under different boundary conditions. We find that the change in the geometry of the system leads to results that are not consistent with results obtained for isotropic two and higher dimension. Bipartite entanglement does not decrease for all RVB dimers due to the different geometry and multiparty entanglement is not formidable. The ramifications of the result have been discussed.

\end{abstract}

\maketitle

\section{Introduction} 

 Resonating valence bond (RVB) states were introduced in 1938 by Pauling 
%to describe quantum mechanical resonance in unsaturated 
%covalent bonds 
in organic, and later in metals and intermetallic, compounds \cite{1}. It was extensively studied in many body physics ever since Anderson, in 1973, presented  the idea of using such states to explain the behavior of Mott insulators \cite{2}. The importance of 
RVB states grew immensely after the proposition of relating such states with high temperature superconductivity \cite{3a, 1a}. 
 %It can be described as a mixture of the singlet pairing of the electrons (with no long range order) at different sites \cite{4a} (see \cite{1a} for 
%detailed description).
%The other applications of RVB states include super conductivity in organic solids and superconductor transitions in diamond doped with Boron. A general description of RVB and its applications is well documented (refer \cite{1a}). 
Investigations on the use of RVB states in quantum information has been initiated in recent years. 
The possibility of topological quantum computation by using such states have been  proposed \cite {2a}, and 
 the entanglement properties of RVB states in isotropic two- and three-dimensional 
%square and higher dimensional 
lattices have also been explored  \cite{5a}. 

In this paper, we consider the RVB state on a ladder, a pseudo two-dimensional system (See Fig. 1). 
\begin{figure}[h!]
\label{fig-chhobi-asol}
\begin{center}
\epsfig{figure= 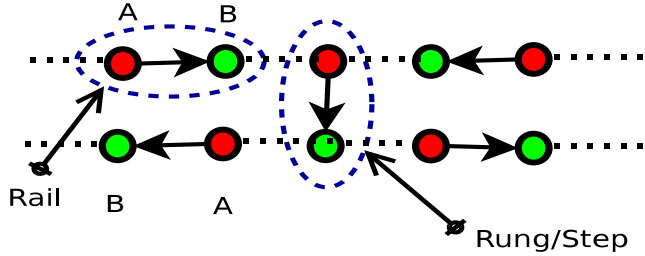, height=.15\textheight,width=0.52\textwidth}
\caption{(Color online.) RVB liquid on a  ladder. Blue  and Red balls  belong to two different sublattices \(A\) and \(B\) respectively. One particular dimer covering is shown. Singlets are always from sublattice A to sublattice B. 
}
\end{center}
\end{figure} 
%Unexpected discovery of high T$_c$ superconductivity in highly doped antiferromagnet ignited interests in low dimensional spin systems \cite {6}. 
Ladder states (obtained by assembling spin chains one next to the other to form \textit{ladders} of increasing width) have recently  generated a lot of  interest in theoretical and experimental condensed matter physics. Ladder states have been proposed in the study of superconductivity in systems like doped  (VO)$_2$P$_2$O$_7$ \cite {5} and the study of transitions from 1-D chains to 2-D spin systems \cite{6}.
It has been found that 
ground states of antiferromagnetic Heisenberg models (with suitable couplings) on ``even ladders'' (consisting of two or more even number of chains) can be  RVB states \cite{6, 7, 8}.
% \textbf{IN WHICH SYSTEM??????????????????????}.
% Hence RVB ladder states are used to study the spin properties of such ladders (for e.g. see \cite{8}). 
%From the quantum information perspective, e.g. 
Quantum information concepts like entanglement, fidelity, of different ladder systems
%, like Heisenberg, random transverse-field Ising model 
 have been studied \cite{Indrani-Igloi}, and protocols for high fidelity transmission of quantum states
through such systems have also been presented \cite{9}. 
%Therefore, the study of bipartite as well as multipartite entanglement  of the RVB ladder can have a applications in condensed matter physics as well as quantum information processing. 

\subsection*{Main results}  We study the entanglement properties of  RVB states on a pseudo-two dimensional ladder   
Precisely, we consider ladders of two chains (\(2\times M\) ladders, N=2M spins), of different lengths. 
We find that the entanglement characteristics of an RVB state on a ladder  
are significantly different from that on an isotropic two or higher dimensional lattice. 
Bipartite entanglement between neighboring sites are of two different types for the case of a ladder. While the bipartite entanglement of the neighboring sites on the rails of the ladder is insignificant, that of the steps (or rungs) is substantial.
 This automatically suggests that genuine multiparty entanglement of the ladder RVB is negligible.
 This is in sharp contrast with the situation on an isotropic 2D or 3D lattice, where the single type of nearest neighbor bipartite entanglement is negligible, 
while the state is genuine multiparty entangled \cite{5a}. We reach our conclusions by numerical simulations as well as via analytical bounds on the entanglement measures. 
The symmetry of the system leads us to use the singlet fraction as a measure of bipartite entanglement, while in the multiparty case, we use  a recently proposed measure called ``generalized geometric measure'' \cite{amadertele, amadernext}.

This change in the entanglement properties 
can be attributed to the change in the geometry of the RVB system: a ladder is not isotropic in all directions. 
%While the general RVB state is isotropic in all directions, a \(2 \times M\) ladder state is not so.  
Our results show that this change in geometry has a marked effect on the entanglement properties of the system, and hence on quantum information tasks possible. 
The results are obtained by using periodic boundary conditions. However, we have also found the similar trend of bipartite entanglement, when there is 
open 
periodic boundary condition. Also the calculations were done for the,
physically interesting, so-called RVB liquid, which we now define. 

\section{RVB liquid} 
%The ladder system under investigation can be thought of two (even) coupled one-dimensional chain and are hence linked by short range spin correlations or RVB interactions \cite{6}. 
The pseudo two-dimensional lattice under consideration can be divided into sub-lattices (see Fig. 1), A and B, in such a way that all the nearest neighbor (NN) sites 
of any site on sub-lattice A belong to sub-lattice B, and vice-versa. Such a  
lattice is called a \textit{bipartite lattice}. Each lattice site on such a bipartite  ladder is occupied by a qubit,
which can for example be the spin state of an electron.
% corresponding to a spin half electron. 
%\begin{figure}[htbp]
%\centering
%\includegraphics[scale=0.40]{file4.png}
%\caption{The 2x4 RVB ladder bipartite lattice, showing A and B lattice sites}
%\end{figure}
For such a system, the  RVB state is defined as 
the equal superposition
%can be written as the superposition 
of all the possible dimer (singlet) coverings that can be formed 
by nearest-neighbor \emph{directed} dimers. Between site \(i\) in sublattice A and site \(j\) in sublattice B, a directed dimer is defined as 
\[
|(i,j)\rangle = \frac{1}{\sqrt{2}}(|\uparrow_i \downarrow_j\rangle - |\downarrow_i \uparrow_j\rangle),
 \]
where \(|\uparrow\rangle\) and \(|\downarrow\rangle\) are respectively the spin-up and spin-down states in the \(z\)-direction, at the corresponding lattice site.
%between the NN(s). 
%Since the NNs surrounding any site of a sub-lattice belongs to the other sub-lattice, 
%dimers can only be formed from a site in A to any one of the neighboring sites in B. 
The (unnormalized) RVB state can then be written as \cite{wv}
\begin{equation}
\label{eq_RVB}
\left|\psi\right\rangle=\sum 
%h(a_1,a_2,...,a_N,b_1,b_2,...b_N)
\left|(a_1,b_1)(a_2,b_2)...(a_N,b_N)\right\rangle,
\end{equation}
where $a_i$ and $b_j$ are  site positions
 and the summation is  over all dimer coverings such that the dimers in every covering satisfy \(a_i \in A\)  and \(b_j \in B\). 
%$(a_i,b_j)$ represents a singlet(dimer) between neighboring sites A and B. For each configuration, every site in the sublattice A, has a unique NN site in the sublattice B 
%to form a dimer. 
Since the dimer coverings that constitute the RVB state are formed by using only  NN dimers, we call the state \(|\psi\rangle\) as an \emph{RVB liquid}.

\section{Bipartite Entanglement of RVB liquid} 

We are now ready to study the behavior of nearest-neighbor bipartite states for an RVB liquid on a ladder. 
%we obtain the wave function as a superposition of all the possible dimers possible in the configuration.  
To obtain the bipartite density matrix, \(\rho_{12}\), between any site (say, 1) in A and one of its nearest neighbors (say, 2) in B, we  take the partial 
trace of the whole RVB liquid over all sites except 1 and 2:
%the states other than sites involved in bipartite entanglement. 
%Hence, it is given by
\begin {equation} 
\rho_{12}=\mbox{Tr}_{\overline{12}} \left|\psi\right\rangle\left\langle\psi\right|
\end{equation} 
where 
%$\rho_{12}$ is the reduced bipartite density matrix and 
$\mbox{Tr}_{\overline{12}}$ represents the partial trace over all sites other 
than \(1\) and \(2\). Rotational invariance of \(|\psi\rangle\), given in Eq. (\ref{eq_RVB}), implies that \(\rho_{12}\) (and all other 
reduced density matrices of \(|\psi\rangle\)) is also rotationally invariant. 
The only rotationally invariant two-qubit states are the singlet and the maximally mixed state. 
Therefore, \(\rho_{12}\) is the 
%Werner state  
%Now,the bipartite density matrix, as per the calculations done is the 
Werner state \cite{ws} 
%(due to the fact that for any two body density matrix, the possible rotationally invariant state is the singlet 
%and the identity matrix,), 
given by
\begin{equation}
\rho_{12}(p) = p\left|(i,j)\right\rangle\left\langle (i,j)\right|+\frac{1-p}{4} I_4,
\end{equation}
where the ``Werner parameter'' \(p\) satisfies $-1/3\leq$p$\leq1$,
 and $I_4$ is the  identity operator of the four-dimensional complex Hilbert space. 
%Note here that the Werner state is entangled for \(p\leq 1/3\) and 
Bipartite entanglement measures, like concurrence \cite{eof}, are  monotonic functions of \(p\). 
Hence, instead of calculating entanglement measures explicitly, we will investigate the behavior of \(p\) with respect to increase of system size. 
Note here that the Werner state is entangled for \(p > 1/3\).
%\textbf{[CITE]!!!}.  

 %he numerical simulation in the next subsection. The bipartite entanglement between two NN sites along the same chain gives us the entanglement for the rail and between NN sites belonging to different chain gives us the entanglement for the rung.
 
%It can be shown that the reduced density matrices of $\left|\psi\right\rangle$ is rotationally invariant and the two body reduced density matrix is a Werner state \cite{5a}. This is due to the fact that for any two body density matrix the possible rotationally invariant state is the singlet and the identity matrix. Hence any bipartite density matrix in general has to be a Werner state, to satisfy the rotational invariance criteria. This has been obtained explicitly for the lower values of N and also holds for higher N. By explicitly obtaining the Werner state we estimate ${p}$ for the system. The BE can then be calculated using \textit {entanglement of formation}\cite{eof}. Hence the BE of the system can be calculated in terms of $\textit{p}$ for systems of increasing size and the extent of change in BE can estimated. From the Werner state it is evident that the bipartite state is maximally entangled for $\textit{p}$=1 and separable for $\textit{p}$=0. The bipartite entanglement between two NN sites along the same chain gives us the entanglement for the rail and between NN sites belonging to different chain gives us the entanglement for the rung.

\subsection{Properties of nearest-neighbor bipartite states for periodic boundary conditions}

In this subsection, we will investigate the bipartite entanglement of two neighboring sites of the ladder 
with periodic boundary conditions. 

\subsubsection{Bipartite Entanglement: Steps versus Rails}

As mentioned earlier, the nearest-neighbor bipartite state of the RVB liquid is a Werner state. 
By symmetry, there are only two types of such bipartite states: along any one of the rails, and on a step. 
We denote the Werner parameter 
\(p\) of a Werner state along a rail as \(p_r\), while that on a step as \(p_s\) (see Fig. 1). 

\vspace{0.6cm}
\begin{figure}[h!]
\label{fig-chhobi-ek}
\begin{center}
\epsfig{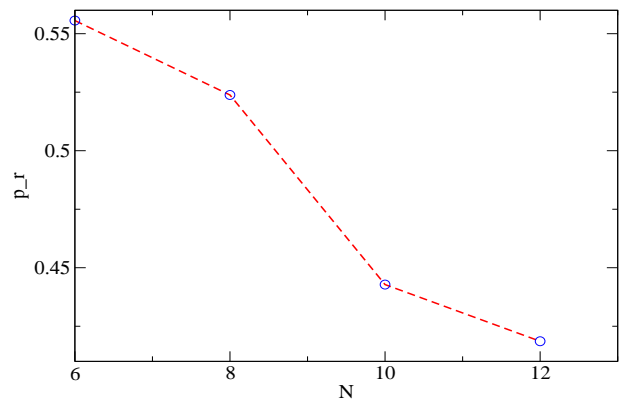}
\caption{(Color online.) Bipartite entanglement on the rails. 
The decrease of the entanglement parameter \(p_r\) with increasing \(N\) is clearly seen.
%for different \(2 \times N\). Here \(N=3, 4, 5, 6\) are plotted. 
}
\end{center}
\end{figure}

As shown in Fig. 2, the values of \(p_r\)  consistently decrease with the increasing system size \(N\), and hence 
 bipartite entanglement on the rails decreases with the increasing \(N\), similar to the case of an isotropic lattice. Interestingly, however, 
the BE of the steps \emph{increases} with respect to \(N\) (see Fig. 3).

\begin{figure}[h!]
\label{fig-chhobi-dui}
\begin{center}
\epsfig{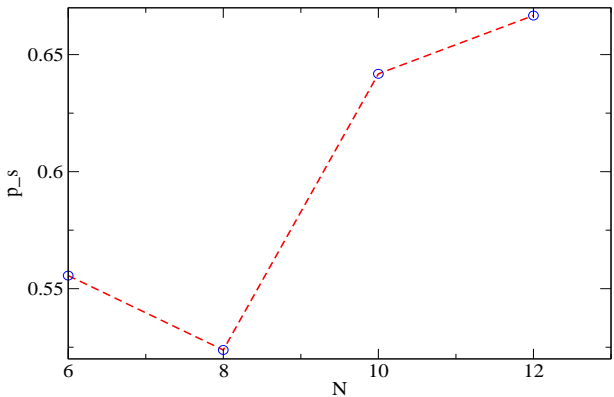}
\caption{(Color online.) Bipartite entanglement on the steps. 
In sharp contrast to the isotropic case, bipartite  entanglement on the steps, as parametrized by \(p_s\),
 \emph{increases} with increasing system size 
%of \(p_s\) along the steps for lattices of different sizes 
\(N\).} 
\end{center}
\end{figure}

Therefore, already at the level 
of bipartite entanglement, one obtains a trade-off between the two different forms of nearest-neighbor entanglement, in the case of 
an RVB liquid on a ladder. Contrast this with the case of a square lattice or any other isotropic
 lattice, where the corresponding RVB liquids have only a single type of nearest-neighbor entanglement, and its value, quite generally, decreases with 
increasing system size \cite{5a}. 
The  numerical evidence presented above for this complementary behavior of the two types of  BE, present in the ladder, 
will later in the paper also be justified by using analytic bounds.

\subsubsection{Regional Entanglement}

To study the behavior of BE between nearest neighbors, 
%To understand the overall bipartite entanglement between NN of the system 
we introduce a quantity, called ``regional entanglement'' (denoted as $p_{avg}$)  for each lattice site.
For every lattice site, there are three NN sites, which form three nearest-neighbor bipartite entanglements with that site. 
(Note that we  are still in the regime of periodic boundary conditions.) 
Regional entanglement at a particular site is defined as the average of these three nearest-neighbor entanglements, as 
quantified by their Werner parameters. 
% and its 3 NN(s), since each site point is entangled to its 3 NN(s)-- two equal rail  and one rung (see Fig. 1).  
Due to periodic boundary conditions, the regional entanglement at each site is the same, so that 
it is a characteristic parameter for a given ladder lattice size. It is a measure of the regional distribution of bipartite 
entanglement at any lattice site.
%It turns out that the amount of entanglement around each site point is same, because due to PBC, the density matrices in the rails and rungs (though different from each other) is same at all sites of the ladder and so is the value of $p$. Hence the average value of $p_{avg}$ attributed to each site is the same for a lattice of a given size. 

The physical significance of $p_{avg}$ can be understood by using the concept of \emph{fidelity of teleportation} \cite{tele, fot}. 
%Let $p_r$ and $p_s$ be the rail and rung values of $\textit{p}$. Now, if one were to c
Suppose that an arbitrary two-dimensional quantum state  is available near a particular lattice site, and 
the task is to quantum teleport \cite{tele} the state to a neighboring lattice site, say the one on the other rail. 
%is randomly teleported to the three nearest neighbors with equal probability. 
The fidelity of teleportation for such an exercise is given by \cite{fot}
\[
F^s_{tele} = \frac{p_s + 1}{2}.
\] 
However, the same protocol that teleports the quantum state to the NN site on the other rail, will also 
teleport the state, with a different teleportation fidelity, to two other NN sites on the same rail \cite{fot}, with the 
new fidelity being 
\[
F^r_{tele} = \frac{p_r + 1}{2}.
\] 
The average fidelity of the transfer is given by 
%. in the case of the RVB ladder, would be given by the relation
\begin{equation}
F_{tele}=\frac{2 F^r_{tele} + F^s_{tele}}{3} = \frac{p_{avg} +1}{2}.
%  \left.\right.\frac{p_r+1}{2}+ \frac{p_s+1}{2},
\end {equation}
%since two of the NN are in a Werner state with $p_r$ (rails) and one with $p_s$ (rung). Hence,
%\begin{eqnarray}
%\label{eq_p_avg}
%F_{tele}&&=
%%\frac{2p_r+ p_s+ 3}{2}
% 3\left.\right. \frac{\left(\frac{2 p_r + p_s}{3}\right)+1}{2}
%= 3\left.\right. \frac{p_{avg} +1}{2}.
%\end {eqnarray}
%%Hence $p_{avg}$ is the value with which the machine would teleport to the three NNs with the same fidelity treating all the sites as the same.

Numerical simulations show that although there are two types of bipartite entanglement, which show complementary 
behavior with increasing system size, the regional entanglement, \(p_{avg}\),
 decreases as the size of the lattice is increased (see Fig. 4). 
%The variation in the average value of $\textit{p}$ is shown graphically below. One can observe that the bipartite entanglement decreases as the size of the quasi two-dimensional system is increased. 
%The lower bound of the decrease can be estimated algebraically considering the individual changes in the values of $\textit{p}$ along the rails and rung.
\vspace{0.7cm}
\begin{figure}[h!]
\label{fig-chhobi-tin}
\begin{center}
\epsfig{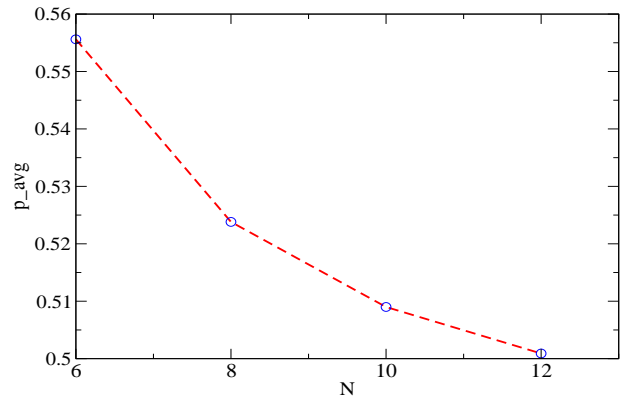}
\caption{(Color online.)  
Regional entanglement in RVB liquids on ladders of different sizes. 
%The decrease of \(p_{avg}\) with respect to different \(2 \times N\) lattices. We consider \(N=3, 4, 5, 6\).
} 
\end{center}
\end{figure}

To sum up, regional entanglement of RVB liquid on ladders is a quantity that mirrors the behavior of nearest-neighbor
bipartite entanglement of isotropic lattice RVBs. However, the internal picture of bipartite entanglement in RVBs
is far richer and depends on the geometry of the lattice. 
%This study only indicates that although the distribution of bipartite entanglement can follow the trade-off of 
%the homogeneous lattice, the internal picture of BE  is  complex and depends on the geometry of the lattice.  
We have also observed that in the case of RVB ladders without periodic boundary conditions, the complementary behavior between bipartite entanglement of rails and steps 
 remains the same  as in the case of ladders with periodic boundary
conditions.

\subsection{Analytical estimates of bipartite entanglement}

\subsubsection{Upper Bounds from Monogamy of Entanglement}

A first estimate on the bipartite entanglement can be obtained by using monogamy of entanglement \cite{moe}. 
Let us consider an arbitrary site \(i_A\), say on sublattice \(A\).
% in the lattice 
If the lattice is finite, and if periodic boundary conditions are not assumed, we assume that \(i_A\) is not on a  boundary step (see Fig. 1). 
Each such site  is surrounded by three NN sites, belonging to the sublattice \(B\).
 Rotational invariance of the state \(|\psi\rangle\)  ensures that  all the three NN bipartite states 
are in Werner states, of which two have Werner parameter \(p_r\), and one has \(p_s\).
The monogamy of entanglement \cite{moe} demands that 
\(2\tau(\rho_r) + \tau(\rho_s) \leq \tau_{1:\small{\mbox{rest}}}(|\psi\rangle)\).
 \(\tau(\rho)\) is the ``tangle'' of the bipartite state \(\rho\), and is defined as
\(\max (0,  \lambda_1 -\lambda_2 - \lambda_3 - \lambda_4)\), where \(\lambda_i\)'s are the square roots of the eigenvalues, in decreasing order,
of \(\rho \tilde{\rho}\). Here, \(\tilde{\rho} = (\sigma_y \otimes \sigma_y) \rho^{*} (\sigma_y \otimes \sigma_y) \), 
where complex conjugation is with respect to the  \(\sigma_z\) eigenbasis \cite{eof,moe}.  
And \(\tau_{1:\small{\mbox{rest}}}(|\psi\rangle)\) is the tangle of the state \(|\psi\rangle\) in any bipartition of one site to rest of the sites,
and is unity.
%Hence t
Also 
%The tangle for each site reads \cite{eof} 
\(\tau(\rho(p_{r/s})) = (3 p_{r/s} -1)^2/4\), 
%\begin {equation}
%\tau(\rho_{AB_k})=(3p-1)^2/4
%\end {equation}
%However the total tangle between the site and its NN cannot exceed 1 and hence
%Therefore, in this case,  the monogamy inequality for a \(N\)-qubit state, \(\sum_{k=2}^N \tau (\rho_{1k}) \leq \tau (rho_{1:\textit{rest}}) \),  is given by 
so that the monogamy inequality reads
\begin{equation}
  (3p_r-1)^2/2 + (3 p_s -1)^2/4 \leq 1,
\end{equation}
%where the rhs is \(1\) due to the fact that \(|psi\rangle\) is maximally entangled with the bipartition 
%\(1:rest\). 
The bound on \(p_r\) and \(p_s\) obtained in the form of the above inequality is depicted in Fig. 5. The 2D projection of Fig. 5 clearly 
shows that the allowed value for \(p_s\) can go up to \(1\) while the same for \(p_r\) is \(\leq 0.8\). Therefore, 
the complementary behavior between the entanglements in the rails and steps can also be anticipated from the monogamy inequality.
\begin{figure}[h!]
\label{fig-chhobi-7}
\begin{center}
\epsfig{figure= 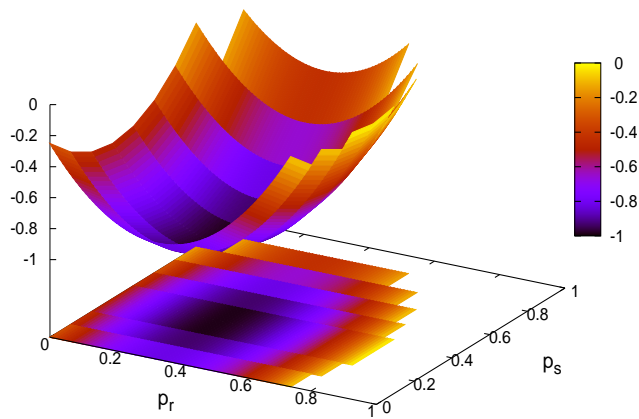, height=.4\textheight,width=0.4\textwidth, , angle = -90}
\caption{(Color online.) Estimate of Werner parameters from the monogamy of entanglement. 
The allowed combinations of \(p_r\) and \(p_s\) are those for which the surface \((3p_r-1)^2/2 + (3 p_s -1)^2/4 - 1\) is negative, and can be 
read off from the projection of the surface. 
}
\end{center}
\end{figure}

\subsubsection{Bounds from Asymmetric Cloning}

Consider again the task of sending an arbitrary auxiliary qubit from any site \(i_A\)  to its three neighboring sites \cite{tele}.
 The teleportation fidelity   of the output state at a site on the same rail (as \(i_A\)) will be \((p_r +1)/2\), while  the fidelity will be \((p_s +1)/2\) 
of the output state on the other rail \cite{fot}.  Quantum mechanics implies that the fidelities  of the output states  in the whole protocol
cannot exceed 
the fidelities  of three approximate clones of the initial state in the optimal \(1 \rightarrow (1+2)\) asymmetric quantum cloning machine \cite{acin, tc}). 
 Therefore, we obtain
\begin{equation}
\label{telebound}
p_r \leq \frac{1}{3}(\sin^2\theta + \sqrt{2} \sin 2 \theta);\quad  p_s \leq 1 - \frac{4}{3} \sin^2 \theta,   
\end{equation}
for all possible \(\theta\). 
Note here that if \(\theta \rightarrow 0\), it would imply \(p_r \leq 0\) and \(p_s \leq 1\). Moreover, 
%a Werner state with 
\(p_r \leq 0\) will imply  
 no bipartite entanglement in the rails while there can be some -- in principle, maximal -- entanglement in the steps. 
Numerical evidence (in Figs. 2 and  3) 
also suggest that this is indeed the trend of \(p_r\) and \(p_s\). In case of  a multipartite state, maximal bipartite entanglement in any of its two-party 
reduced density matrices  
leads to no genuine multipartite entanglement. 
The question remains whether \(\theta\) will be zero for large system-size.
% \(N\), in our system. 

To obtain estimates of \(\theta\),  we insert the values of \(p_r\) and \(p_s\), that have been obtained from 
the numerical simulations (for different \(N\)) in Sec. IIIA, in the inequalities in (\ref{telebound}).
  We then solve the above inequalities to obtain \(\theta \in S_1^N\) for the first inequality for a fixed \(N\), and 
similarly \(\theta \in S_2^N\). 
We now consider the allowed \(\theta\) lying in the intersection \(S_1^N \cap S_2^N\), and plot 
\(\theta_{\max} = \max\{\theta: \theta \in S_1^N \cap S_2^N\}\) with respect to \(N\).
%\begin{figure}[htbp]
%\centering
%\includegraphics[scale=0.40]{shade.JPG}
%\caption{The 2x4 RVB ladder bipartite lattice, showing A and B lattice sites}
%\end{figure}
%ties, instead of inequalities and obtain two values of \(\theta\). 
%%\textbf{We had chosen the MAXIMUM of \(theta\) from the region \(S_1^N \cup S_2^N\), that is we took the UNION. 
%%Do u remember why we did so? why did we not choose the intersection which would give a more stringent bound?
%%If u think that the intersection cannot be chosen, then please write to me why that is so. If u think that intersection IS allowed, then please send me the 
%%new numbers. We can then change the figure.}  
%We now choose the maximum \(\theta_{\max}\) among two \(\theta\)s.  
\begin{figure}[h!]
\label{fig-chhobi-8}
\begin{center}
\epsfig{figure= 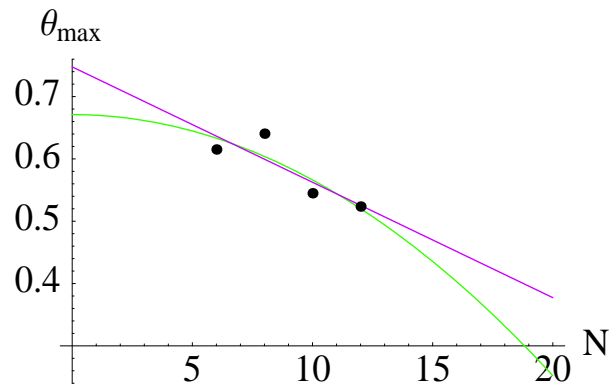, height=.22\textheight,width=0.45\textwidth}
\caption{(Color online.) 
Decrease of \(\theta_{\max}\) for increasing \(N\). The dots represent \(\theta_{\max}\) obtained from 
Eq.  (\ref{telebound}) corresponding to \(N = 6, 8, 10, 12\). Pink and green lines are respectively linear (\( 0.747664 - 0.0185155 x\)) and 
quadratic  (\(0.671077 -0.0010471 x^2\)) fits of the points. The corresponding mean square errors are 
%for quadratic and   
\(1.22 \times 10^{-3}\) for the linear fit and \(1.06 \times 10^{-3}\) for the quadratic. 
}
\end{center}
\end{figure}
We find that \(\theta_{\max}\) is decreasing with the increase of the size of the lattice, as shown in  Fig. 6. 
%The decrease of 
%\(\theta_{\max}\) is  quadratic with \(N\). 
The inequalities in Eq. (\ref{telebound}) also shows  the complementarity between the entanglement of rails and steps (see Fig. 7). 
\begin{figure}[h!]
\label{fig-chhobi-eta}
\begin{center}
\epsfig{figure= 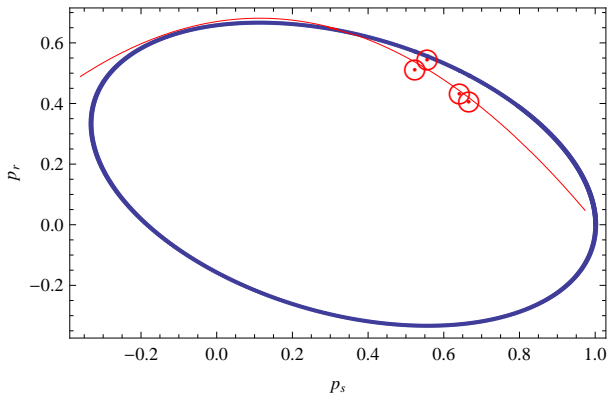, height=.22\textheight,width=0.45\textwidth}
\caption{(Color online.) Complementary behavior of the entanglement on the rail with that on the step. 
The bound on \(p_r\) and \(p_s\), as obtained from the Eq. (\ref{telebound}), for all possible values of \(\theta\), is depicted by the ellipse in the figure.
The allowed combinations of \(p_r\) and \(p_s\) fall inside the ellipse. 
The (red) open curve is the quadratic fit \((-0.858 x^2 + 0.241 x + 0.67)\) of the (red) circles, which are in turn 
the values found by exact calculations for \(N=6,8,10,12\).
%calculated  by using Eq. (\ref{telebound}).
 All calculated points are within the bounding ellipse. It is evident that as \(p_s\) attains the maximum value 1, the value of \(p_r\) goes to 0.
}
\end{center}
\end{figure}

 This gives us  further evidence 
%(from the realtion of \(\theta\) with multipartite entanglement) 
that there is  negligible or no  multipartite entanglement for large lattice size,  
in contrast to the case of isotropic lattices \cite{5a}. 
We will concretize this evidence by calculating a measure of genuine 
multipartite entanglement for RVB liquids on ladders, in the next section.

\section{Negligible Genuine Multipartite Entanglement}

The numerical and analytical studies on nearest-neighbor bipartite entanglement in the preceding section already suggest 
that the RVB liquid on a ladder has negligible or no genuine multiparty entanglement. This is due to the evidence presented 
that the entanglement of the states on the steps are near-maximal or maximal. A maximally entangled state in \(d \otimes d\) must be 
pure \cite{unpub} (cf. \cite{Garisto-Hardy}), and hence cannot have any correlation, classical or quantum,  with any other 
quantum system (cf. \cite{Artur}). 

%The behavior of bipartite entanglement  already suggests that there can not be a significant genuine multipartite entanglement. 
To quantify the amount of genuine multiparty entanglement present in the RVB liquid on ladders of different sizes,
%show the behavior of multipartite entanglement, 
we consider a recently introduced genuine multipartite entanglement measure
%, introduced recently 
called generalized geometric measure (GGM) \cite{amadertele, amadernext}.
%, and is given by
The GGM of  an \(N\)-party pure quantum state \(|\phi_N\rangle\) is defined as
\begin{equation}
{\cal E} ( |\phi_N\rangle ) = 1 - \Lambda^2_{\max} (|\phi_N\rangle ), 
\end{equation}
where  \(\Lambda_{\max} (|\phi_N\rangle ) =
\max | \langle \chi|\phi_N\rangle |\), with  the maximization being over all pure states \(|\chi\rangle\)
that are not genuinely \(N\)-party entangled. 
It was shown in Ref. \cite{amadernext} that 
\begin{equation}
\label{label}
{\cal E} (|\phi_N \rangle ) =  1 - \max \{\lambda^2_{ A: B} |  A \cup  B = \{1,2,\ldots, N\},  A \cap  B = \emptyset\}.
 \end{equation}
%1 - \max\{e_{i:\textit{rest}}^{max} , e^{max}_{ij:\textit{rest}} \cdots |i \neq j\} , 
%\end{equation}
 where \(\lambda_{A:B}\) is  the maximal Schmidt coefficients in the \(A: B\) bipartite split  of \(|\phi_N \rangle\) .

In Fig. 8, we find that indeed the genuine multipartite entanglement measure \({\cal E}\) for 
the RVB liquid on a ladder, decreases with increasing \(N\).  
We also notice that when performing the 
maximization in  Eq. (\ref{label}) for obtaining the GGM, 
the maximum Schmidt coefficient is obtained when the maximum number of steps is included on one side of the
bipartition. This can be explained by the complementary behavior  
of bipartite entanglement in steps and rails as discussed  earlier. Therefore, the trend of multipartite as well as bipartite entanglements 
indicate that for large \(N\), only bipartite entanglement in the steps will remain,  while bipartite entanglement of the rails as well as multipartite entanglement 
of the whole RVB liquid will disappear.  
\vspace{1cm}
\begin{figure}[h!]
\label{fig-chhobi-9}
\begin{center}
\epsfig{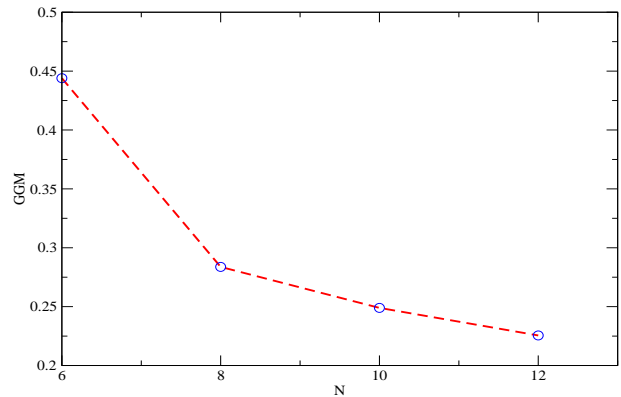}
\caption{(Color online.) Genuine multipartite entanglement measure of RVB liquid on ladders. 
%\({\cal E}\) with respect to different \( 2 \times N\) system. \(N = 3, 4, 5, 6\). 
The figure clearly shows that the genuine multipartite entanglement decreases with the increase of system size \(N\). }
\end{center}
\end{figure}

%GGM FIGURE WILL  BE HERE

\section{Conclusion}
We have considered the resonating valence bond liquid on a ladder  with periodic boundary conditions. 
We have found two different kinds of bipartite entanglement: While the bipartite entanglement on the steps 
%the rung of the ladder is 
is increasing, that of the rails is  decreasing, with the increase of the size of the lattice. Both numerical and analytical bounds 
support  this thesis. 
% Contrary to the isotropic lattices like square or cubic lattice, 
Moreover, genuine multipartite entanglement  of the ladder decreases with increasing system size.
%also decreases while substantial bipartite entanglement in the rung is present. 
This is in sharp contrast with the situation in isotropic lattices, where  
%Therefore, the 
same state (RVB  liquid) has negligible bipartite entanglement, but substantial multipartite entanglement.
%can show a contrasting picture of bipartite as 
%well as multipartite entanglement, with the change of  geometry. 
This shows that geometry can play an important role in determining the entanglement properties of multiparty quantum states.

\acknowledgments

We thank Ujjwal Sen for  helpful discussions.

\end{document}